\documentclass[journal]{IEEEtran}
\ifCLASSINFOpdf
\usepackage[pdftex]{graphicx}
\DeclareGraphicsExtensions{.pdf,.jpeg,.png}
\else
\usepackage[dvips]{graphicx}
\DeclareGraphicsExtensions{.eps}
\fi
\usepackage[cmex10]{amsmath}
\usepackage{nicefrac}
\usepackage{hyperref}
\hypersetup{colorlinks=true} 
\usepackage{tabularx}

\usepackage{algorithmic}
\ifCLASSOPTIONcompsoc
  \usepackage[caption=false,font=normalsize,labelfont=sf,textfont=sf]{subfig}
\else
  \usepackage[caption=false,font=footnotesize]{subfig}
\fi
\usepackage{fixltx2e}

\usepackage{stfloats}
\usepackage{url}

\begin{document}
%
\title{On Higher Order Positive Differential Energy Operator}
%
%
%
\author{Amirhossein~Javaheri\IEEEauthorrefmark{1},
        Mohammad~Bagher~Shamsollahi,~\IEEEmembership{Senior Member,~IEEE}
\thanks{\IEEEauthorrefmark{1}Amirhossein~Javaheri and Mohammad~Bagher~Shamsollahi are both with Sharif~University~of~Technology, Tehran, Iran. (email:~javaheri\_amirhossein@ee.sharif.edu; email: mbshams@sharif.edu)}}
\maketitle

\begin{abstract}
\boldmath
The higher order differential energy operator (DEO), denoted via $\Upsilon_k(x)$, is an extension to the second order famous Teager-Kaiser operator. The DEO helps measuring the higher order gauge of energy of a signal which is useful for AM-FM demodulation. However, the energy criterion defined by the DEO is not compliant with the presumption of positivity of energy. In this paper we introduce a higher order operator called Positive Differential Energy Operator (PDEO). This operator which can be obtained using alternative recursive relations, resolves the energy sign problem. The simulations demonstrate that the proposed operator can outperform DEOs in terms of Average Signal to Error Ratio (ASER) in AM/FM demodulation. 
\end{abstract}


%
\IEEEpeerreviewmaketitle

\section{Introduction}
\label{sec:Introduction}
%
%
%
%

\IEEEPARstart{T}{he}-continuous-time DEO of order $k$ denoted via $\Upsilon_k(x)$ is defined as follows \cite{1}:
\begin{equation}
\label{eq_1}
\Upsilon_k(x)\equiv[x,x^{(k-1)}]=\dot{x}x^{(k-1)}-xx^{(k)},   k=0,±1,±2,\ldots
\end{equation}
where $[,]$ denotes the Lie bracket and $x^{(k)}$ denotes the $k$th-order derivative of $x(t)$. The DEOs are generally applied in algorithms which help demodulate AM-FM signals also known as Energy Separation Algorithms (ESA) \cite{2,2_2}.
These algorithms are used in several applications in biomedical and communications signal processing \cite{3,4,5,6,7}. The continuous-time Teager-Kaiser Energy Operator (TKEO) \cite{8,9,10} also denoted by $\Psi(x)$ is a special case of DEO with $k=2$. There is also a recursive relation between a continuous-time DEO of order $k$ and DEOs of preceding orders \cite{1}:
\begin{equation}
\label{eq_2}
\Upsilon_{k}(x(t))=\frac{d}{dt}\Upsilon_{k-1}(x(t))-\Upsilon_{k-2}(\dot{x}(t))
\end{equation}
The discrete-time DEO is obtained by discretizing the continuous-time definition of $\Upsilon_{k}(x)$. There are several approaches to discretization of \eqref{eq_1} as proposed in \cite{1}. One approach is to first discretize the Lie bracket and then replace derivatives with respective time shifts (ordinary differences), which yields: 
\begin{equation}
\label{eq_5}
\Upsilon_k(x[n])=x[n]x[n+k-2]-x[n-1]x[n+k-1]
\end{equation}
The resulting operator is a discrete DEO which hereafter we call it discrete ordinary DEO\footnote{Same notation is used in this paper for both continuous and discrete-time ordinary operators obtained via simple discrete time shift}. Another choice is to replace the $m$th-order derivatives in \eqref{eq_1} with backward ~$x[n]-x[n-~1]$
\IEEEpubidadjcol
 or symmetric $1/2(x[n+1]-x[n-1])$ differences of same order, where $x[n]=x(nT_s)$.
As with the continuous-time derivative, the $m$th order discrete difference is obtained by consecutively applying the first order difference $m$ times. i.e.:
\begin{equation}
\label{eq_6}
\Delta_t^{(m)}(x)\equiv \Delta_t(\Delta_t^{(m-1)}(x))
\end{equation}
where $t$ denotes the type of difference (backward or symmetric). With $\Delta_b$ representing the first-order backward difference we have:
\begin{equation}
\label{eq_7}
\Delta_b\equiv \Delta_b^{(1)}(x) = x[n]-x[n-1]=\Delta_b[n]*x[n]
\end{equation}
In fact $\Delta_b$ is a linear operator with $\Delta_b(z)=(1-z^{-1})$, the $Z$ transform of its impulse response  $\Delta_b[n]$. Thus for backward difference of higher order we have $\Delta_b^{(m)}(z)=(1-z^{-1})^m$.
Hence, applying the inverse $Z$ transform we get:
\begin{equation}
\label{eq_8}
\Delta_b^{(m)}(x[n])=\sum_{j=0}^{m}{(-1)^j\binom{m}{j}x[n-j]}
\end{equation}
Similar approach for symmetric difference of order $m$ yields:
\begin{equation}
\label{eq_9}
\Delta_s^{(m)}(x[n])=\frac{1}{2^m} \sum_{j=0}^{m}{(-1)^j\binom{m}{j}x[n+m-2j]}
\end{equation}
Now replacing $x^{(m)}(t)$ with backward or symmetric discrete difference of order $m$, the discrete symmetric  and backward DEOs are obtained \cite{1}. For backward DEO denoted by $\Upsilon_{k_b}$ we have:
\begin{equation}
\label{eq_10}
\Upsilon_{k_b}(x_n)=\Delta_b(x_n)\Delta_b^{(k-1)}(x_n)-x_n\Delta_b^{(k)}(x_n)
\end{equation}
Also for symmetric DEO ($\Upsilon_{k_s}$) we get:
\begin{equation}
\label{eq_11}
\Upsilon_{k_s}(x_n)=\Delta_s(x_n)\Delta_s^{(k-1)}(x_n)-x_n\Delta_s^{(k)}(x_n)
\end{equation}
Now using the definition in \eqref{eq_8}, we can restate \eqref{eq_10} in terms of ordinary DEOs: 
\begin{equation}
\label{eq_12}
\Upsilon_{k_b}(x_n)=\sum_{j=1}^{k}{(-1)^j \binom{k-1}{j-1}\Upsilon_{j}(x_{n-j+1})}
\end{equation}
Generalizing the definition of \eqref{eq_5} to account for negative values of $k$, we may write: 
\begin{equation}
\label{eq_13}
\Upsilon_{k}(x_n)=-\Upsilon_{2-k}(x_{n+k-1}), \qquad k<0
\end{equation}
Thus for symmetric DEO it can be shown that: 
\begin{IEEEeqnarray}{rCl}
\label{eq_14}
 \Upsilon_{k_s}(x_n) &=&\frac{1}{2^k} \Bigg\{ \sum_{j=0}^{k-1} (-1)^j \binom{k-1}{j} \Big( \Upsilon_{k-2j}(x_n) \nonumber \\
&& + \Upsilon_{k-2j}(x_{n+1}) \Big) \Bigg\}
\end{IEEEeqnarray}
The DEOs are generally used to help demodulate AM-FM signals like $x(t)=a(t)\cos(\phi(t))$. Therefore it is important for DEOs to output as much simplified combination of instantaneous amplitude and frequency of the signal as possible. The simplest AM-FM signal is a simple cosine. The output of continuous-time DEO to this signal is \cite{1}: 
\begin{equation}
\label{eq_15}
\setlength{\nulldelimiterspace}{0pt}
\Upsilon_k (A \cos (\omega t+\theta) )=\left\{\begin{IEEEeqnarraybox}[\relax][c]{l's}
0,& $k=1,3,5,\ldots$\\
(-1)^{1+\frac{k}{2}} A^2 \omega ^k,& $k=2,4,6\ldots$%
\end{IEEEeqnarraybox}\right.
\end{equation}
For the discrete counterpart $x[n]=A \cos (\Omega n+\theta)$, discrete ordinary DEO yields \cite{1}: 
\begin{equation}
\label{eq_16}
D_{k}(A,\Omega) :=\Upsilon_k (A \cos (\Omega n+\theta) )=A^2 \sin(\Omega) \sin [(k-1)\Omega]
\end{equation}
Using the expansions \eqref{eq_8} and \eqref{eq_9} the output of discrete backward and symmetric DEOs to $x[n]$ would also be obtained as: 
\begin{align}
\label{eq_17}
& D_{k_b}(A,\Omega) := \Upsilon_{k_b}(A \cos (\Omega n+\theta)) =\nonumber \\ 
& \left\{ \!\!\!\!
\begin{array}{ll} 
2^k (-1)^{\frac{k+1}{2}} A^2 \sin^k(\frac{\Omega}{2})\dfrac{\sin\left( (\frac{k}{2}-1) \Omega\right)+\sin\left(\frac{k}{2}\Omega\right)}{2}\! &\\
\hfill k=1,3,5,\ldots &\\
2^k (-1)^{1+\frac{k}{2}} A^2 \sin^k(\frac{\Omega}{2})\dfrac{\cos\left( (\frac{k}{2}-1) \Omega\right)+\cos\left(\frac{k}{2}\Omega\right)}{2}\! &\\
\hfill k=2,4,6\ldots &\\
\end{array} \right. 
\end{align}
And
\begin{align}
\label{eq_18}
\setlength{\nulldelimiterspace}{0pt}
 D_{k_s}(A,\Omega) &:=\Upsilon_{k_s} (A \cos (\Omega n+\theta) ) \nonumber \\ 
&= \left\{\begin{IEEEeqnarraybox}[\relax][c]{l's}
0,& $k=1,3,5,\ldots$\\
(-1)^{1+\frac{k}{2}} A^2 \sin ^k (\Omega),& $k=2,4,6\ldots$%
\end{IEEEeqnarraybox}\right.
\end{align}
The discrete symmetric DEO is preferred to discrete ordinary and backward DEOs in the sense that it yields identically zero output for odd $k$ and is much more similar to the output of continuous-time DEO \eqref{eq_15}. But there is still another approach to discretize continuous DEO as proposed in \cite{1} which is referred to as discrete alternative DEO. 
This approach is based on shifting the center point time location of the sample windows where discrete differences used to approximate \eqref{eq_1} are defined upon. In fact the aim of alternative discretization is to equalize the center point of sampling windows for discrete approximation of both terms $\dot{x}x^{(k-1)}$ and $xx^{(k)}$
(For more details refer to \cite{1}). Thus the definition of discrete alternative DEO is as follows: 
\begin{align}
\label{eq_19}
& \Upsilon_{k_a}(x_n) =\nonumber \\ 
& \left\{ \!\!\!\!
\begin{array}{ll} 
\Delta_b(x_n)\Delta_b^{(k-1)}(x_{n+\frac{k-1}{2}})-x_n\Delta_b^{(k)}(x_{n+\frac{k-1}{2}}) \! &k=3,5,\ldots \\
\Delta_b(x_{n+1})\Delta_b^{(k-1)}(x_{n-1+\frac{k}{2}})-x_n\Delta_b^{(k)}(x_{n+\frac{k}{2}}) \! &k=2,4,\ldots \\
\end{array} \right. 
\end{align}
It can be easily shown that the alternative DEO satisfies the recursive relation below for $k>2$:
\begin{align}
\label{eq_20}
& \Upsilon_{k_a}(x_n) =\nonumber \\ 
& \left\{ \!\!\!\!
\begin{array}{ll} 
\Delta_b(\Upsilon_{{k-1}_a}(x_n))- \Upsilon_{{k-2}_a}(\Delta_b(x_{n+1})) \! &k=3,5,\ldots \\
\Delta_b(\Upsilon_{{k-1}_a}(x_{n+1}))- \Upsilon_{{k-2}_a}(\Delta_b(x_{n})) \! &k=2,4,\ldots \\
\end{array} \right. 
\end{align}
with the initial conditions $\Upsilon_{1_a}(x_n) = 0$ and $\Upsilon_{2_a}(x_n) = \Psi(x_n)$. 
Using equation \eqref{eq_8} the output of the alternative DEO for a discrete cosine would also be acquired as:
\begin{align}
\label{eq_22}
\setlength{\nulldelimiterspace}{0pt}
&D_{k_a}(A,\Omega) := \Upsilon_{k_a} (A \cos (\Omega n+\theta) )=\nonumber \\
&\left\{\begin{IEEEeqnarraybox}[\relax][c]{l's}
0,& $k=3,5,\ldots$\\
(-1)^{1+\frac{k}{2}} A^2 \sin ^k \Big(\frac{\Omega}{2}\Big) \cos ^2 \Big(\frac{\Omega}{2}\Big),& $k=2,4,\ldots$%
\end{IEEEeqnarraybox}\right.
\end{align}
\section{Positive DEO}
\label{sec:PositiveDEO}
We can obtain other types of DEOs by considering \eqref{eq_2} in other alternative discrete forms. For instance if we discard the separate cases of even and odd $k$ in recursive definition of the alternative DEO \eqref{eq_20} and consider the general case: 
\begin{equation}
\label{eq_23}
\Upsilon_{k_{a'}}(x_n)= \Delta_b(\Upsilon_{{k-1}_{a'}}(x_n))-\Upsilon_{{k-2}_{a'}}(\Delta_b (x_n))
\end{equation}
we obtain a new operator which may be formulated as: 
\begin{equation}
\label{eq_24}
\Upsilon_{k_{a'}}(x_n)= \sum_{i=0}^{\lfloor \frac{k}{2}\rfloor}(-1)^i \binom{k-2-i}{i}\Delta_b^{(k-2-2i)}\Big(\Psi \big(\Delta_b^{(i)}(x_n)\big)\Big)
\end{equation}
The output of this operator to $x[n]=A \cos(\Omega n +\theta)$ is:
\begin{align}
\label{eq_25}
\setlength{\nulldelimiterspace}{0pt}
&D_{k_{a'}}(A,\Omega) :=\Upsilon_{k_{a'}} (A \cos (\Omega n+\theta) )=\nonumber \\
& \left\{\begin{IEEEeqnarraybox}[\relax][c]{l's}
0,& $k=3,5,\ldots$\\
(-1)^{1+\frac{k}{2}} \Psi \Bigg( \Delta_b^{(\frac{k}{2}-1)} \Big(A \cos (\Omega n+\theta) \Big) \Bigg),& $k=2,4,\ldots$%
\end{IEEEeqnarraybox}\right.
\end{align}
Inspiring from this relation, we can introduce a new operator which we call it Positive Differential Energy Operator or PDEO. The term \emph{positive} emphasizes the positivity of total energy measured via this operator. The total energy $E_t$ of a signal  $x(t)$ is defined to be the integral of energy operator output of the signal $O(x)$, over time domain, i.e.,
\begin{equation}
\label{eq_26}
E_t = \int_{-\infty}^{\infty} {O(x(t)) dt}
\end{equation}
For the popular Squared Energy Operator (SEO), we can state the total energy in terms of energy spectral density of the signal:
\begin{equation}
\label{eq_27}
E_t = \int_{-\infty}^{\infty} {x^2(t) dt}=\frac{1}{2\pi} \int_{-\infty}^{\infty} {\vert X(j\omega) \vert^2 d\omega}
\end{equation}
where we have assumed $x(t)$ is a real-valued square integrable signal. This criterion always yields a positive value for the energy of the signal. However, for DEO defined via \eqref{eq_1} we may obtain a negative value of total energy for some $k$. For example the DEO-based total energy of a real continuous-time signal $x(t)$ is:
\begin{equation}
\label{eq_28} 
 \int_{-\infty}^{\infty} \Upsilon_{k}(x(t)) = (-1)^{1+\lfloor \frac{k}{2} \rfloor}\frac{1}{\pi} \int_{-\infty}^{\infty} {\vert X(j\omega) \vert^2 \omega^k d\omega}
 \end{equation}
which always yields a negative value for $k=4l$. The problem of energy sign also appears for discrete-time DEOs assuming that the total energy of a discrete-time signal is computed via $E_n=\sum_{n=-\infty}^{\infty}{O(x[n])}$. For ordinary discrete-time DEO we have: 
\begin{align}
\label{eq_29} 
\sum_{n=-\infty}^{\infty}{\Upsilon_{k}(x[n])}&= \frac{1}{\pi} \int_{-\pi}^{\pi}  D_k (\vert X(e^{j\Omega}) \vert,\Omega) d\Omega \nonumber \\
&= \frac{1}{\pi} \int_{-\pi}^{\pi}  \vert X(e^{j\Omega}) \vert^2 \sin [(k-1)\Omega] \sin (\Omega) d\Omega
 \end{align}
 which gives $-\frac{1}{4}$ for $X(e^{j\Omega})=j\sin (\Omega)$ and $k=4$. In general for discrete DEOs it can be shown that: 
 \begin{equation}
 \label{eq_29b}
\sum_{n=-\infty}^{\infty}{\Upsilon_{k_t}(x[n])}= \frac{1}{\pi} \int_{-\pi}^{\pi}  D_{k_t} (\vert X(e^{j\Omega}) \vert,\Omega) d\Omega 
 \end{equation}
where $t$ denotes the type of DEO (backward, symmetric or alternative) and $D_{k_t}$ denotes the (desired) output to a simple cosine as obtained in \eqref{eq_17}, \eqref{eq_18} and \eqref{eq_22}. For example for symmetric DEO we may write ($x[n]$ is real):
\begin{equation}
\label{eq_31}
\sum_{n=-\infty}^{\infty}{\Upsilon_{k_s}(x[n])}=(-1)^{1+\lfloor \frac{k}{2} \rfloor} \frac{1}{\pi} \int_{-\pi}^{\pi}  \vert X(e^{j\Omega}) \vert^2 \sin^k (\Omega) d\Omega
 \end{equation}
which is always negative for $k=4l$. Since a negative value of energy is not physically meaningful, it is required to define a new operator for accurately measuring the higher order energy of a signal. This operator should always yield a positive value for the total energy and the proposed operator, PDEO, does so. The PDEO of order $k$ for a continuous-time real-valued signal is defined as: 
\begin{equation}
\label{eq_32}
\setlength{\nulldelimiterspace}{0pt}
\Upsilon_{k_p}(x)=\left\{\begin{IEEEeqnarraybox}[\relax][c]{l's}
\Psi\left(x^{(m_k)}\right), & $k=1,3,5,\ldots$\\
\Upsilon_{-1} \left(x^{(m_k)}\right), & $k=2,4,6\ldots$%
\end{IEEEeqnarraybox}\right.
\end{equation}
where $\Upsilon_{k_p}(x)$ denotes the $k$th-order PDEO and $x^{(m_i)}$ denotes the $m_i$th-order continuous-time derivative. The value of $m_k$ is chosen such that a PDEO of order $k$ measures the $k$th-order energy or $k$th-order spectral moment of energy of a signal. Hence:
\begin{equation}
\label{eq_33}
\setlength{\nulldelimiterspace}{0pt}
m_k=\left\{\begin{IEEEeqnarraybox}[\relax][c]{l's}
\dfrac{k-1}{2}, & $k=1,3,5,\ldots$\\
\frac{k}{2}-1, & $k=2,4,6\ldots$
\end{IEEEeqnarraybox}\right.
\end{equation}
Now if we measure the total energy of a real-valued signal based on PDEO, we will get
\begin{equation}
\label{eq_34}
\int_{-\infty}^{\infty}{\Upsilon_{k_p}(x(t)) dt}=
\frac{1}{\pi}  \int_{-\infty}^{\infty}{\omega^k \vert X(j\omega) \vert^2 d\omega} \
\end{equation}
The advantage of \eqref{eq_34} over \eqref{eq_28} is that for even $k$, the integrand is an even function of $\omega$ and thus $E_t$ is always positive (since we have assumed $x(t)$ is real and consequently $\vert X(j\omega) \vert $ is real and even \cite{11}). This implies that the PDEO based total energy is always of non-negative sign. We may have another approach to \eqref{eq_32} by considering the recursive relation \eqref{eq_2} and eliminating the term $\frac{d}{dt} (\Upsilon_{k-1}(x))$. We consequently obtain an operator similar to \eqref{eq_32} for the case of even  $k$. In fact since the output of DEO for a simple cosine is considered to be either zero (for odd $k$) or a constant function of $A$ and $\Omega$, its time derivative is identically zero and thus we can discard $\frac{d}{dt} (\Upsilon_{k-1}(x))$ from \eqref{eq_2} and only write:
\begin{equation}
\label{eq_35}
\Upsilon_{k_{p'}}(x)=-\Upsilon_{{k-2}_{p'}}(\dot{x})
\end{equation}
when applied for even values of $k$ this recursive relation with initial condition $\Upsilon_{2_{p'}}(x)= \Psi(x)$, yields:
\begin{equation}
\label{eq_36}
\setlength{\nulldelimiterspace}{0pt}
\Upsilon_{k_{p'}}(x) =\left\{\begin{IEEEeqnarraybox}[\relax][c]{l's}
0 & $k=1,3,5,\ldots$\\
(-1)^{1+\frac{k}{2}} \Psi (x^{(m_2)})  & $k=2,4,6,\ldots$
\end{IEEEeqnarraybox}\right.
\end{equation}
which only differs in sign from what we defined in \eqref{eq_32} for even values $k$.
\section{Discrete PDEO}
\label{sec:Discrete PDEO}
Similar approach to what was explained in section \ref{sec:Introduction} for discretization of continuous-time DEO can be applied to discretize \eqref{eq_32}. If we use simple time-shift (or ordinary difference) to approximate discrete-time derivate, then for $m$th-order continuous-time derivative, the discrete approximation would be $x^{(m)}(t)\approx x_n-x_{n-m}$.
Substituting this in \eqref{eq_32} and using \eqref{eq_13} we get: 
\begin{align}
\label{eq_37}
\Psi (x_n-x_{n-m}) = & \Psi(x_n)+\Psi(x_{n-m})-\Upsilon_{-m+2}(x_n) \nonumber \\
& -\Upsilon_{m+2}(x_{n-m})
\end{align}
Substituting $m$ from \eqref{eq_33} yields the ordinary discrete PDEO: 
\begin{align}
\label{eq_38}
& \Upsilon_{k_p}(x_n)=\nonumber \\ 
& \left\{ \!\!\!\!
\begin{array}{ll} 
\Psi(x_n)+\Psi(x_{n-\frac{k-1}{2}})-\Upsilon_{\frac{-k+5}{2}}(x_n) - \Upsilon_{\frac{k+3}{2}}(x_{n-\frac{k-1}{2}})\! &\\
\hfill k=1,3,5,\ldots &\\
\Psi(x_n)+\Psi(x_{n-\frac{k}{2}+1})-\Upsilon_{-\frac{k}{2}+3}(x_n) - \Upsilon_{\frac{k}{2}+1}(x_{n-\frac{k}{2}+1}) \! &\\
\hfill k=2,4,6,\ldots &\\
\end{array} \right. 
\end{align}
Note that $\Upsilon_{1_p} (x_n)\equiv \Upsilon_{2_p} (x_n)\equiv 0$. The symmetric or backward difference approximation of \eqref{eq_32} would yield quite a more complex expression. To symmetrically approximate the continuous-time derivative, we take use of \eqref{eq_9}. Thus we have:
\begin{align}
\label{eq_42}
& \Upsilon_{k_{ps}}(x)  \equiv \Psi(\Delta_s^{(m_k)}(x_n)) =\nonumber \\ 
&\frac{1}{2^{2m_k}}  \sum_{i=0}^{m_k} \sum_{j=0}^{m_k} \binom{m_k}{i} \binom{m_k}{j} (-1)^{i+j} x_{n-m_k-2i} x_{n-m_k-2j} - \nonumber \\
&\frac{1}{2^{2m_k}} \sum_{i=0}^{m_k} \sum_{j=0}^{m_k} \binom{m_k}{i} \binom{m_k}{j} (-1)^{i+j} x_{n-1+m_k-2i} x_{n+1+m_k-2j} 
\end{align}
where $\Upsilon_{k_{ps}}(x) $ denotes the symmetric PDEO of order $k$ and it should be noted that $m_k$ is a function of $k$ as in \eqref{eq_33}. This long expression can be restated in terms of discrete ordinary DEOs in a more compact form:
\begin{align}
\label{eq_43}
 \Upsilon_{k_{ps}}(x) &= \nonumber \\
& \frac{1}{2^{2m_k}}  \sum_{i=0}^{m_k} \sum_{j=0}^{m_k} \binom{m_k}{i} \binom{m_k}{j} (-1)^{i+j}  \Upsilon_{2(i+j-1)}(x_{n+m_k-2j}) 
\end{align}
The same procedure can also be applied to obtain the mathematical expression for backward PDEO. i.e.:
\begin{align}
\label{eq_44}
 \Upsilon_{k_{pb}}(x) & \equiv \Psi(\Delta_b^{(m_k)}(x_n)) \nonumber \\
& = \sum_{i=0}^{m_k} \sum_{j=0}^{m_k} \binom{m_k}{i} \binom{m_k}{j} (-1)^{i+j}  \Upsilon_{(i-j+2)}(x_{n-i}) 
\end{align}
Furthermore it can be shown that the total energy of a discrete signal based on discrete PDEOs satisfies equation \eqref{eq_29b} where the desired terms $D_{k_{p}}$, $D_{k_{pb}}$, $D_{k_{ps}}$ for ordinary, backward and symmetric PDEOs respectively, are given as follows:
\begin{align}
\label{eq_41}
D_{k_p} (A,\Omega) = & \Upsilon_{k_p}(A \cos (\Omega n+ \theta) )=\nonumber \\ 
& \left\{ \!\!\!\!
\begin{array}{ll} 
4 A^2  \sin^2(\Omega) \sin^2 \Big(\dfrac{k-1}{4}\Omega \Big) \! & k=1,3,5,\ldots \\
4 A^2 \sin^2(\Omega) \sin^2 \Big(\dfrac{k-2}{4}\Omega \Big) \! & k=2,4,6,\ldots
\end{array} \right. 
\end{align}
\begin{align}
\label{eq_46}
D_{k_{pb}} (A,\Omega) =& \Upsilon_{k_{pb}}(A \cos (\Omega n+ \theta))=\nonumber \\ 
& \left\{ \!\!\!\!
\begin{array}{ll} 
2^{k-1} A^2 \sin^2(\Omega) \sin^{k-1}\!\left(\frac{\Omega}{2}\right) \! & k=1,3,5,\ldots \\
2^{k-2} A^2 \sin^2(\Omega) \sin^{k-2}\!\left(\frac{\Omega}{2}\right) \! & k=2,4,6,\ldots 
\end{array} \right. 
\end{align}
\begin{align}
\label{eq_45}
D_{k_{ps}} (A,\Omega) =& \Upsilon_{k_{ps}}(A \cos (\Omega n+ \theta) ) =\nonumber \\ 
& \left\{ \!\!\!\!
\begin{array}{ll} 
A^2  \sin^{k+1}(\Omega) \! & k=1,3,5,\ldots \\
A^2  \sin^k(\Omega) \! & k=2,4,6,\ldots 
\end{array} \right. 
\end{align}
It is clear that the desired terms above are even functions of $\Omega$ and hence the PDEO-based total energy is positive.
\section{Numerical Error Analysis}
\label{sec:NumericalError}
\begin{figure*}[!t]
\centering
\subfloat[$k=4$]{\includegraphics[width=2.5in]{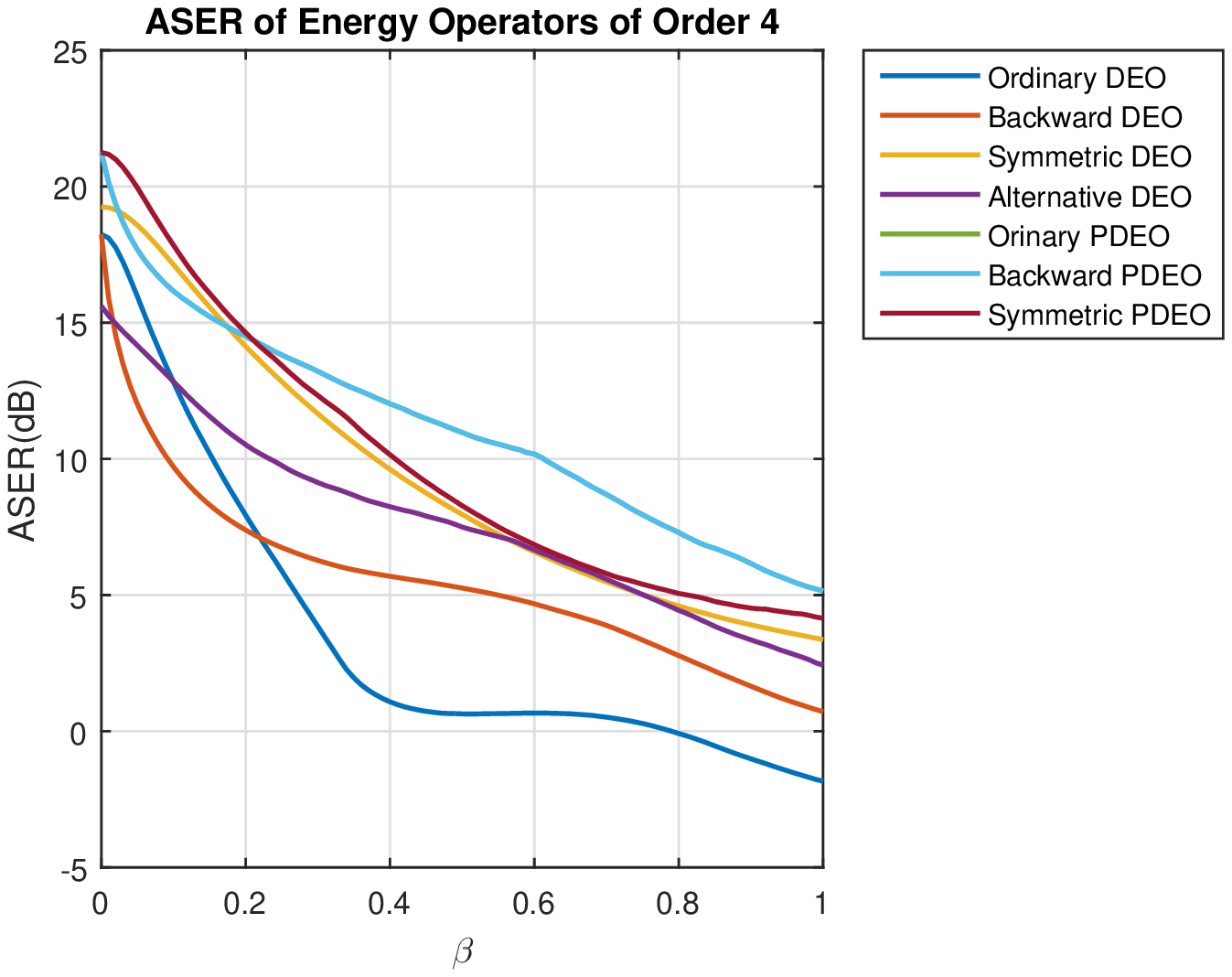}%
\label{fig1:a}}
\subfloat[$k=6$]{\includegraphics[width=2.5in]{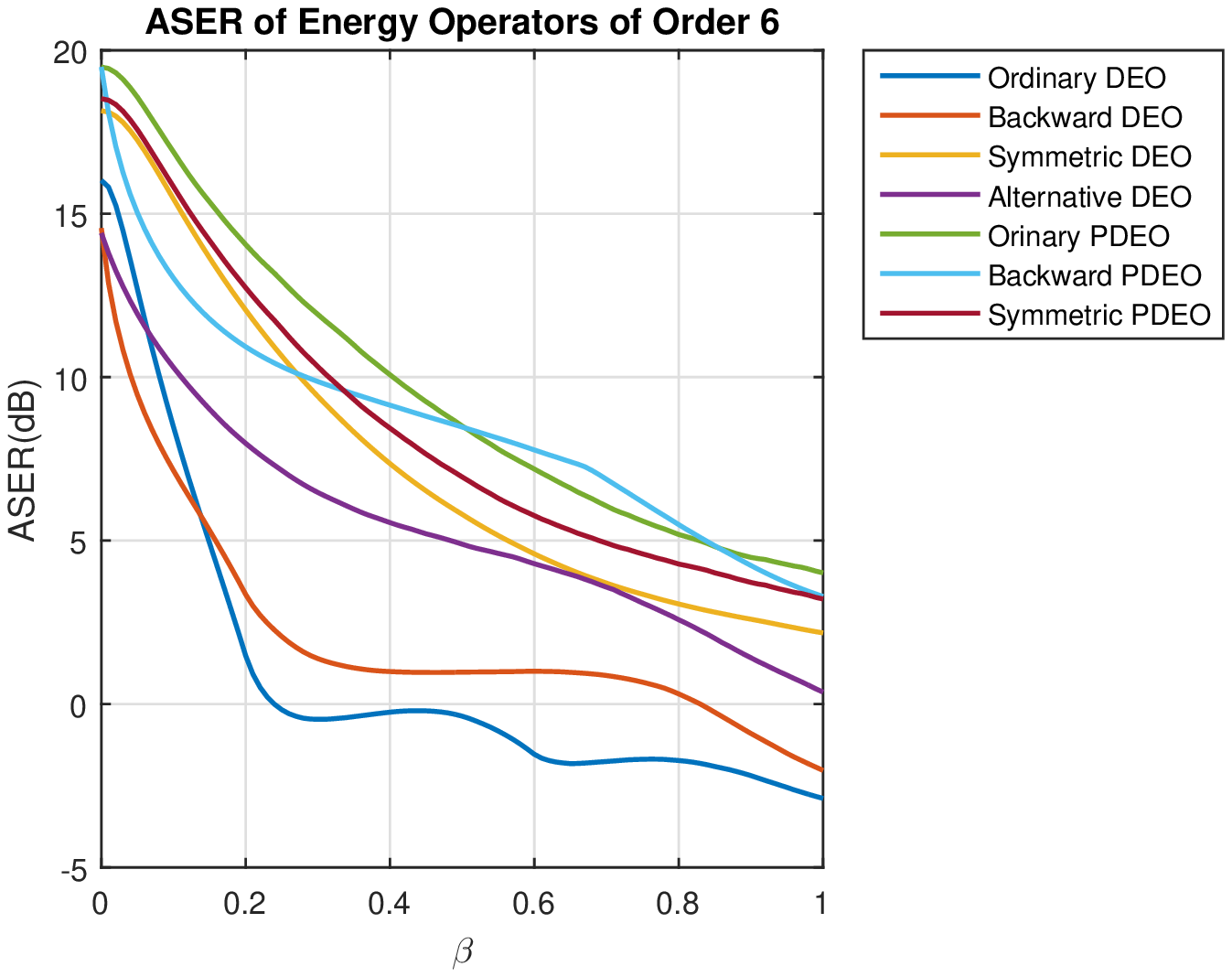}%
\label{fig1:b}} 
\subfloat[$k=8$]{\includegraphics[width=2.5in]{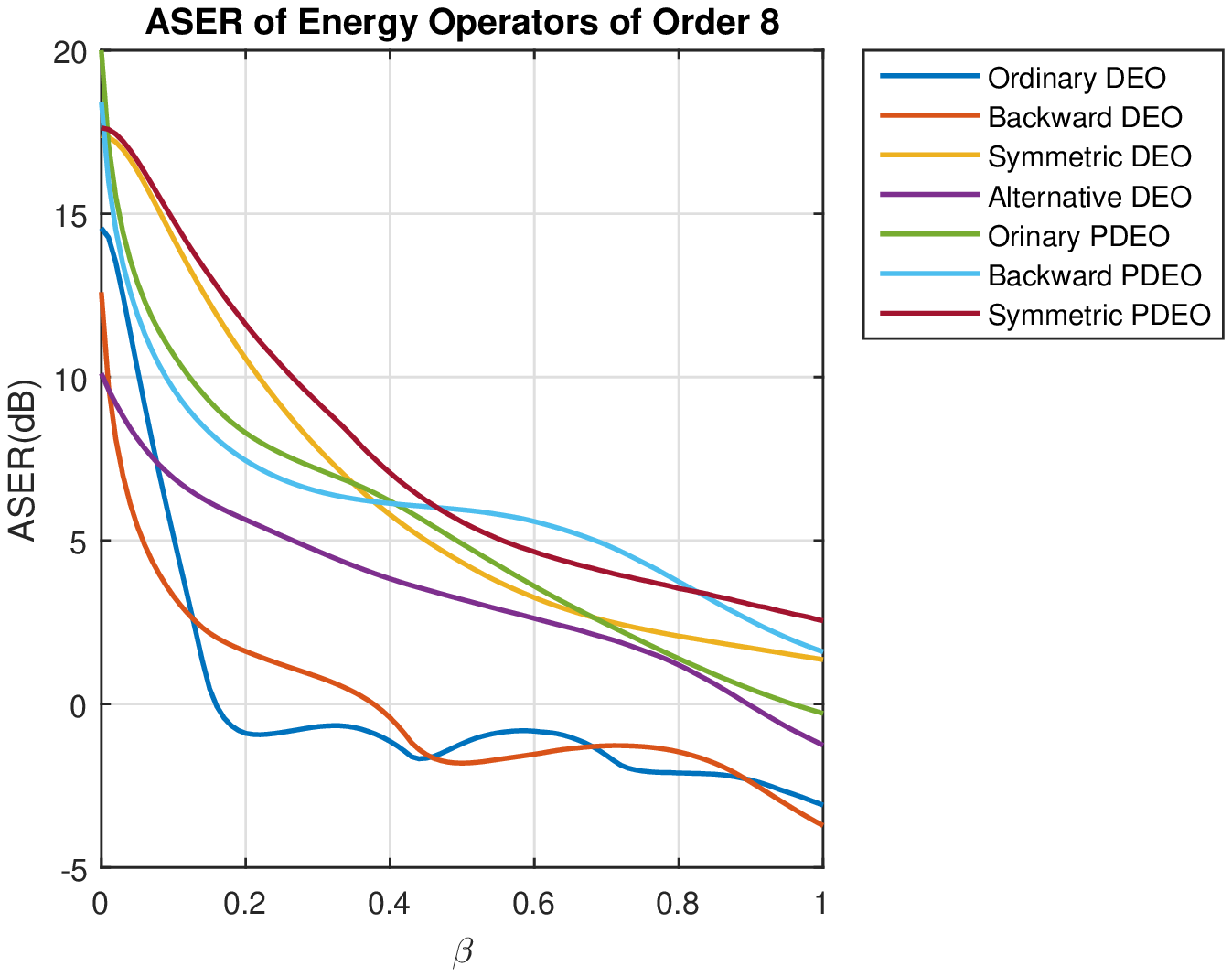}%
\label{fig1:c}}
\caption{ASER performance of discrete DEOs and PDEOs verse $\beta$ for FM modulation \eqref{eq_50} with $\Omega_c=\frac{\pi}{2}$. }
\label{fig_1}
\end{figure*}

\begin{figure*}[!t]
\centering
\subfloat[$k=4$]{\includegraphics[width=2.5in]{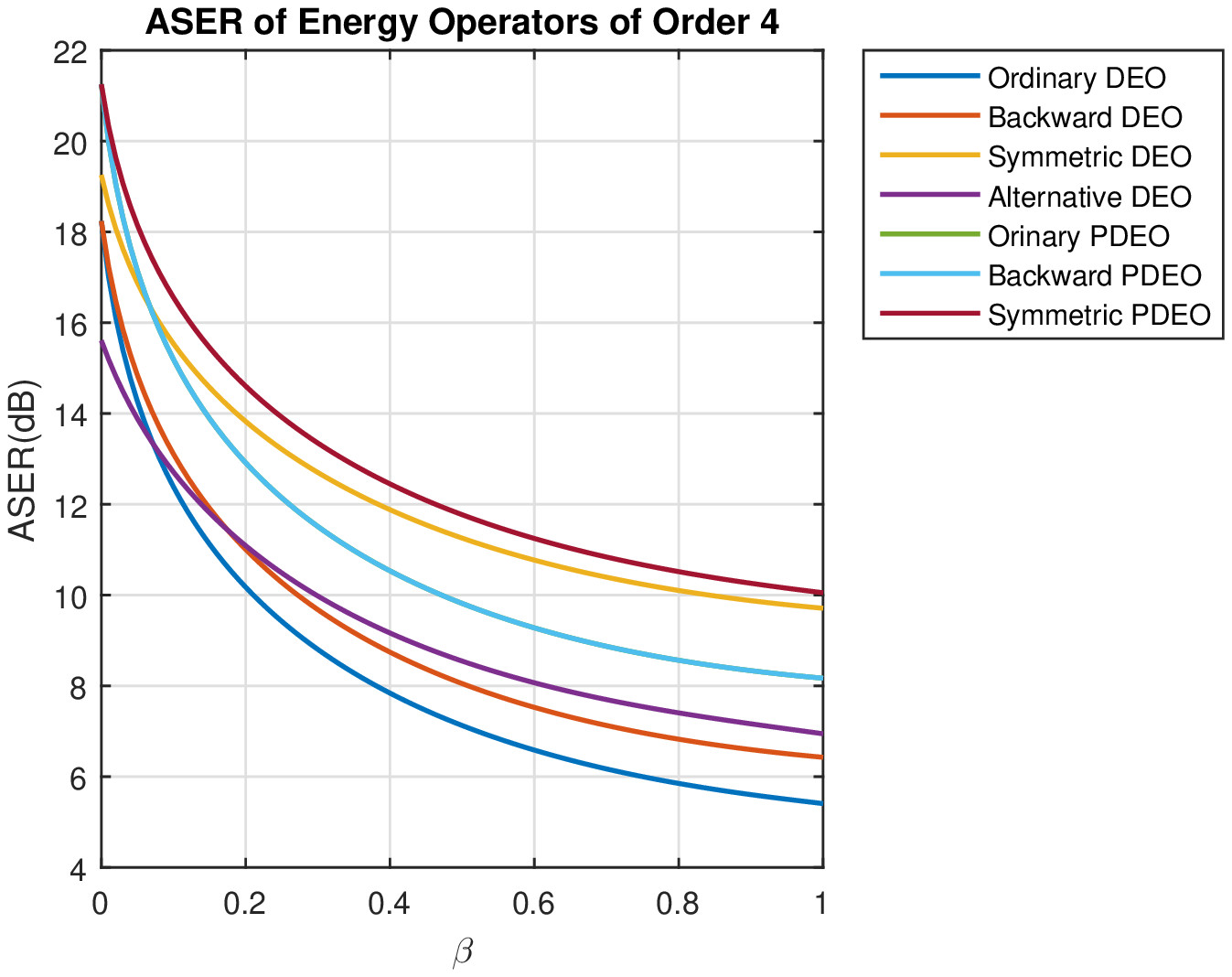}%
\label{fig2:a}}
\subfloat[$k=6$]{\includegraphics[width=2.5in]{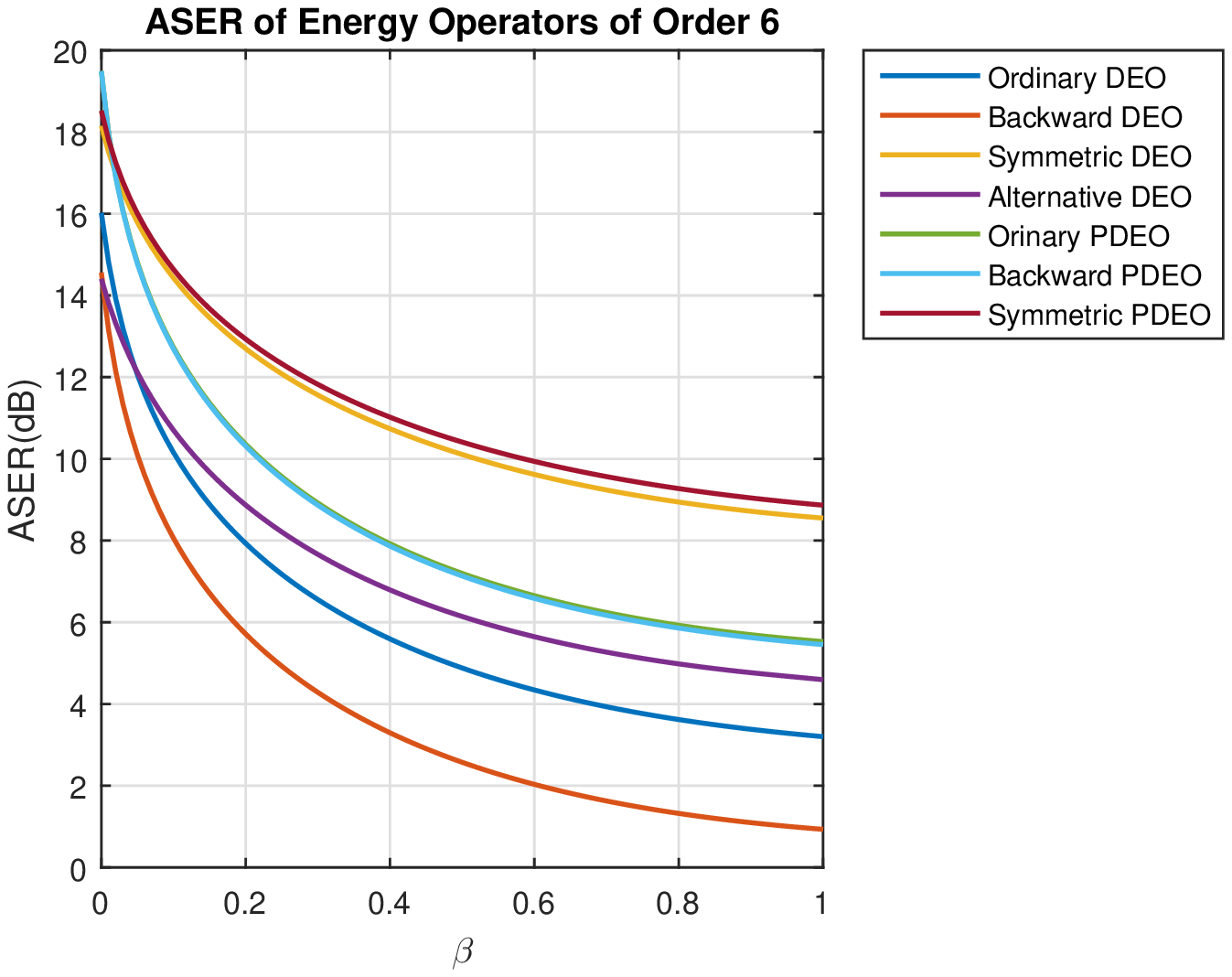}%
\label{fig2:b}}
\subfloat[$k=8$]{\includegraphics[width=2.5in]{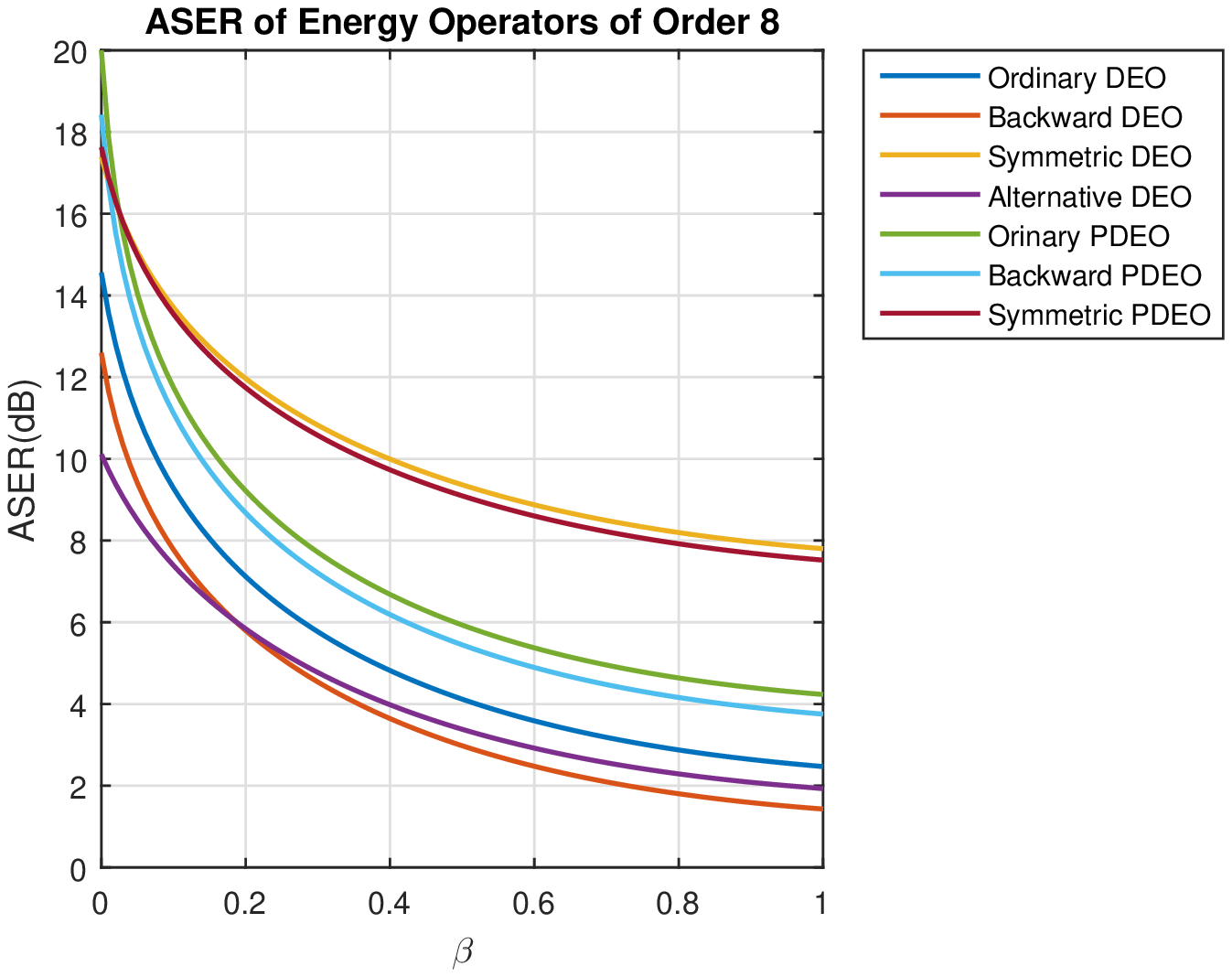}%
\label{fig2:c}}
\caption{ASER performance of discrete DEOs and PDEOs verse $\beta$ for AM modulation \eqref{eq_51} with $\Omega_c=\frac{\pi}{2}$.}
\label{fig_2}
\end{figure*}

In this section we compare the performance of different types of discrete DEOs we have already studied based on the error of approximation. The approximation error is defined (as in \cite{12}) to be the undesired terms available in the output of the energy operator to an AM-FM input signal while the desired term refers to the multiplication of instantaneous amplitude and frequency. This definition of course is valid for second-order operator, but we can extend it to higher order operators as well. In fact the output approximation error of a higher order operator is defined as the difference between the output of the operator to the AM-FM signal $x[n]=A[n] \cos⁡(\phi[n])$ and the output to a simple cosine $x[n]=A \cos⁡(\Omega n+\theta)$ where $A$ and $\Omega$ are subsequently substituted with $A[n]$ and $\Omega[n]$ and $\Omega[n]\equiv \frac{d}{dn}(\phi[n])$. For example the approximation error of a $4\mathrm{th}$-order backward DEO to $x[n]=A[n] \cos⁡(\sum_n \Omega[n] )$ is:
\begin{equation}
\label{eq_47}
E[n]=\Upsilon_{4_b} (x[n])-D[n]
\end{equation}
where $D[n]=D_{4_b}(A,\Omega)$ denotes the desired output specified by \eqref{eq_17}.
We use ASER or Average Signal to Error Ratio as the performance criterion which is defined as \cite{12}:
\begin{equation}
\label{eq_49}
\mathrm{ASER}=\dfrac{\vert D \vert _{\mathrm{ave}}}{\vert E \vert _{\mathrm{ave}}}
\end{equation}
For comparison we consider two separate classes of AM and FM signals. The FM signals are produced as:
\begin{equation}
\label{eq_50}
x_{\mathrm{FM}} [n]_{\lambda ,\beta} = A \cos \left( \sum_n \Omega_i [n] \right)\!, \quad \Omega_i [n]=\Omega_c (1+\beta q[n])
\end{equation}
where $\Omega_c$ is the carrier frequency, $\beta$ is the variable denoting the ratio of FM modulation, i.e., $\beta={\Omega_\Delta}/{\Omega_c}$, and $q[n]$ is a normalized message signal. We assume $q[n]$ is a cosine in the form of $q[n]=\cos⁡(\lambda \Omega_c+\theta)$ where $\lambda $ is the ratio of message signal bandwidth to the carrier frequency. The AM signals are also produced according to:
\begin{equation}
\label{eq_51}
x_{\mathrm{AM}} [n]_{\lambda ,\beta} = A_i [n] \cos \left( \Omega_c n+\theta \right)\!, \quad A_i [n]=A (1+\beta q[n])
\end{equation}
We consider combinations of AM and FM signals in which both $\lambda$ and $\beta$ vary. We choose $\lambda \in (0\!:\!0.02\!:\!0.5)$ and $\beta \in (0\!:\!0.01\!:\!1)$. For each value of $\beta$ we average the ASER over $\lambda$ and then we plot the ASER in dB versus $\beta$. Fig. \ref{fig_1} and \ref{fig_2} show the ASER performance of DEOs and PDEOs of different orders for $\Omega_c=\frac{\pi}{2}$. As depicted in Fig. \ref{fig_1}, discrete symmetric PDEO outperforms discrete symmetric DEO and mostly achieves the highest ASER for both AM and FM demodulation for $\beta \approx 1$. It is also obvious that ordinary and backward PDEOs have better performance compared to their DEO counterparts, especially for large values of $\beta$ in AM/FM demodulation.
\section{Conclusion}
\label{sec:Conclusion}
In this paper we introduced a higher order energy operator called PDEO which resolves the energy sign problem and yields positive value of total energy for all $k$. We showed that this operator may also be derived using both the definition of the discrete alternative DEO and the continuous-time recursive relation \eqref{eq_2} discarding the term corresponding to odd order operators. We obtained relations for both the continuous and the discrete-time definitions of PDEO. We used two approaches to discretize PDEO using backward and symmetric differences to approximate continuous-time derivatives. We also obtained formulas for the output of discrete DEOs and PDEOs to simple periodic signals. The simulation results showed that the proposed PDEO can outperform higher order discrete DEOs in terms of average signal to error ratio. This helps developing energy separation algorithms applicable for more accurate amplitude and frequency extraction which can also be more robust to noise 

\ifCLASSOPTIONcaptionsoff
  \newpage
\fi

\end{document}